# Metalearning-Informed Competence in Children: Implications for Responsible Brain-Inspired Artificial Intelligence

Singh[a], Chaitanya

**Abstract**: This paper offers a novel conceptual framework comprising four essential cognitive mechanisms that operate concurrently and collaboratively to enable metalearning (knowledge and regulation of learning) strategy implementation in young children. A roadmap incorporating the core mechanisms and the associated strategies is presented as an explanation of the developing brain's remarkable cross-context learning competence. The tetrad of fundamental complementary processes is chosen to collectively represent the bare-bones metalearning architecture that can be extended to artificial intelligence (AI) systems emulating brain-like learning and problem-solving skills. Utilizing the metalearning-enabled young mind as a model for brain-inspired computing, this work further discusses important implications for morally grounded AI.

## 1. Introduction

Children display impressive learning abilities (Shrager & Siegler, 1998), being endowed with a brain that represents one of the most sophisticated learning systems on Earth. Language learning, together with the related spheres of language acquisition and speech discrimination, is one of the many accomplishments of humans in early childhood. Infants quickly learn to discriminate verbalizations regardless of their native linguistic environment (Gualtieri & Finn, 2022). Most children, early on in their lives, can address the problem of mapping sound to language with relative ease. Young children also demonstrate a remarkable ability to develop abstractions based on insight and visual acuity, which enables them to recognize facial and emotional cues efficiently (Woodard et al., 2022). In yet other encounters with sensory stimuli, early childhood learners commonly demonstrate the prowess to generalize from an individual phenomenal experience (Watson, 2019). How is it that these children can admirably capture the underlying relationships across acoustic, linguistic, visual, and other contexts to solve complex learning challenges without repeated exposure to domain-specific inputs? What mechanisms in their brain might enable these astonishing feats?

The human ability to perceive high-level information in input data to form rule-based abstractions and other meaningful representations necessitates some basic learning underpinnings. Learning, whether general-purpose or domain-specific, is a quintessential aspect of cognitive development and performance. It embodies the process of gaining new knowledge (NK) and other attributes (Gross, 2015)—namely, behaviors, perceptions, preferences, principles, and proficiencies (B4P)—which improve our current knowledge, intuitive reasoning, and critical understanding of the world (Maass et al., 2023). Research shows that learning can be driven by multiple modalities, including experiences (Gualtieri & Finn, 2022), observations and patterning (Garrigan et al., 2018), and rewards (Eckstein et al., 2021). Experience-driven learning or *crystalized intelligence* helps to boost the adaptability of semantic networks in a developing brain (Gualtieri & Finn, 2022). Garrigan et al. (2018) recognize that learning acquired via observations and patterning promotes the regulation of emotion, and that even 4-month-olds can differentiate between emotions from facial expressions. Learning from rewards, or value-based outcome prediction and choice selection, which is known as reinforcement learning or RL (Eckstein et al., 2021), facilitates our understanding of heuristics (e.g., trial-and-error learning) and decision-making (Niv et al., 2015). RL is observed in simple scenarios of Pavlovian and operant conditioning as well as in intricate settings involving metalearning and multi-context learning (Eckstein et al., 2021). In the RL context, learning is theorized as the development of relationships among actions, effects, and motivations. These relationships help drive the decision-making process in scenarios where comparable motivations exist (Niv et al., 2015).

---

[a] University of California, Berkeley.

Deep and meaningful learning, especially in the formative years, is crucial to building robust learning futures that support a greater potential for new learning, effective skills development, subsequent scholastic achievement (Curran & Kitchin, 2019; Kang et al., 2019; Le et al., 2019; Watts et al., 2014; Watts et al., 2018), and extra-scholastic success (Zhang & Zhang, 2019). Innovation in the artificial intelligence (AI) domain has emerged from emulating human neurobiological capacities (Kim et al., 2021; Mehonic & Kenyon, 2022), especially, the ability of humans to adapt their learning (i.e., NK-B4P) in novel contexts (Alam, 2021). However, advancements in artificial neural nets, which are fueling the explosion of AI solutions across the new digital transformation space, have generally progressed unguided by human moral and ethical[1] behaviors and concerns. Accordingly, a deeper understanding of factors contributing to the efficiency, efficacy, and ethicality of the human learning machine, particularly, in its development stages, holds the promise to shape the future of responsible smart innovation. The human brain is wired for thinking and learning, forming cognition about one's own thinking and learning, and operationalizing motivational beliefs and attitudes to influence the application and development of cognition and metacognition. Self-regulated learning (SRL) theory posits that the interactions among cognition, metacognition, and motivation guide human learning (Schraw et al., 2006). Therefore, the SRL construct may be considered as a learning model that integrates diverse cognitive, metacognitive, and motivational elements (Berger, 2023). SRL embodies the human capacity for dynamically understanding and regulating individual learning enterprises that are directed and restricted by personal goals and environment characteristics (Schraw et al., 2006; Berger, 2023). Accordingly, related research generally interprets SRL as being a superset of metacognition (Berger, 2023). Furthermore, Cao et al. (2022) acknowledge the centrality of SRL in fostering deep-seated and purposeful learning. By linking SRL with the cognitive mechanisms (*enablers*) of attention, explore-exploit tradeoff, feedback, and transfer, this paper proposes a novel metacognitive framework of metalearning to serve two objectives. One, help explain the highly effective as well as morally congruent learning outcomes generated by young children. Two, draw essential implications to inform future technological advancements toward responsible AI.

The proposed explicatory framework focuses on a parsimonious set of complementary enabling mechanisms and a narrowly defined target audience (which are respectively designated herein as the *process* and *population* dimensions of a learning model) with learning experiences across a broadly characterized landscape of epistemic challenges (denoted as *province*). Moreover, by examining the mechanics of learning to learn within the scaffolding of metalearning-enabling mechanisms, this paper seeks to identify the framework's core characteristics which support young children's morally driven learning successes. The discerned framework attributes form the basis for drawing key implications for charting a path forward toward responsible AI innovation. Other models and frameworks—including those advanced by Shrager and Siegler (1998), Rule et al. (2020), Garrigan et al. (2018), and Drigas et al. (2023)—of human learning and cognitive development, by and large, offer process-population-province combinations that diverge from the unique configuration (and the attendant set of attributes) embraced by the framework proposed in this paper. Shrager and Siegler (1998) formulate a unified computational model, integrating metacognitive and associative processes, of young children's strategic development in the use of simple strategies for arithmetic addition tasks. Rule et al. (2020) posit a broad framework supporting the domain-agnostic synthesis, implementation, and examination of knowledge as representations akin to computer code to yield formal learning processes that seek to generalize learning across a vast tapestry of human problem contexts. Garrigan et al. (2018) assimilate various complex processes and components underlying diverse theories and perspectives from developmental psychology and social neuroscience into a wide-ranging framework of moral development and decision-making. Drigas et al. (2023) attempt to unify assorted knowledge, including theoretical components, from several evidence-based studies on human cognition and learning into a nine-layered hierarchical framework of metalearning that is presented as a self-directed and continuous learning model. Whereas Drigas et al. (2023), Garrigan et al. (2018), and Rule et al. (2020) consider the process space to be expansive, contributing significantly to model complexity, Shrager and Siegler (1998) narrowly characterize the learning province and add to the challenge of applying their computational model across domains. In contrast, this paper integrates a bare-bones selection of complementary cognitive processes into a cohesive conceptual metalearning framework with implications for shaping responsible AI. As an important aspect of SRL, metalearning encompasses a higher-order learning awareness that enables the developing brain to learn effectively and adaptively in diverse situations. Essentially, learning competence across contexts[2] is shaped and guided by metalearning component strategies, namely, those that effectuate attentional control for inducing metacognitive awareness, explore-exploit harmonization to foster learning capacity, feedback orchestration for monitoring and evaluation, and learning transfer to direct and incrementally adjust NK-B4P.

## 2. Metalearning

Drigas et al. (2023) define metalearning as comprising perceptual *meta-processes* that enable learners to be deliberate architects and managers of their own learning models. Metalearning, also referred to as *metacognitive learning* in this paper, captures the cognitive and metacognitive processes engaged in acquiring new knowledge, improving current understanding, upgrading skills, and revising beliefs and behaviors. It subsumes learning to learn[3], the ability or fundamental awareness to reflect on one's own learning and learning processes. A metalearning-enabled brain possesses the competence to adapt and optimize its learning strategies, selectively drawing on previous knowledge to foster cognitive and metacognitive processes. Cognitive processes involve meaning-making strategies, including analysis, synthesis, conceptualization, decision-making, and problem-solving. Metacognitive processes, in contrast, incorporate activities and operations for dynamically monitoring and directing one's own cognitive processes as well as the knowledge about these processes and one's own self (Berger, 2023; Schraw et al., 2006). The conventional SRL representation includes three key elements: cognition, metacognition, and motivation (Schraw et al., 2006). SRL is described as the human enterprise that focuses on actively and autonomously strategizing, regulating, and assessing individual learning and the associated learning processes so much so that it influences individual behaviors, motivations, and metacognitive skills (Cao et al., 2022). Notwithstanding the epistemic notion of SRL, the research in this field overwhelmingly depicts SRL representations as emphasizing process over knowledge (Berger, 2023). In this paper, we posit a streamlined framework which educes a variant of the classical SRL triadic characterization (Schraw et al., 2006) from an irreducible set of mechanisms: attention, explore-exploit tradeoff, feedback, and transfer. The resultant cohesive scaffolding seeks to harmonize process and knowledge for high-performance learning among young children. The procedures, functions, and activities that the developing brain is motivated to perform under each mechanism are subsumed in the repertoire of strategies at its disposal. This work formalizes the proposed parsimonious blueprint as the metalearning framework which elicits an adaptation of the traditional SRL representation, with the adapted SRL triad comprising learning, metalearning, and motivation.

There exists an extensive corpus of knowledge on SRL and metacognition (Schraw, 1998, 2006; Berger, 2023). An assortment of research studies characterizes the latter as a construct (Berger, 2009), notion (Drigas, 2023; Flavell, 1979), phenomenon (Schraw, 1998), model (Prather et al., 2020; Zhang & Zhang, 2019), or a set of general skills (Schraw, 1998), expounding its importance in improving learning outcomes (Berger, 2009; Deng et al., 2011; Drigas, 20023; Khasawneh et al., 2020; Zhang & Zhang, 2019). In contrast, metacognitive learning as a distinctive domain does not seem to enjoy the same prominence across literary works. Even less apparent in literature is a consistent theoretical or empirical narrative on the characteristic composition of metalearning. Furthermore, it is unclear whether this composition varies between student populations with and without learning disabilities, especially in the presence of interactive and adaptive AI technology-based mediation. Biggs and Telfer (1987) suggest that the occurrence of metalearning coincides with the development of awareness among learners about their active involvement in the learning process. As with metacognition conceptualized in Schraw et al. (2006) and Berger (2023), this paper incorporates two related traditional mainstays or dimensions (i.e., knowledge and regulation) into the conception of metalearning, albeit with adaptations. A variation along the first time-honored classical dimension, which casts the learning agents' knowledge as their awareness about their attention-mediated understanding of the given context, is included under the attention mechanism. The accommodation along the second dimension considers the regulation of learning and the associated learning strategies as being served by the substructure of feedback in collaboration with the attention mechanism. Besides the individual adaptations of the conventional metacognition-based components, this paper proposes two other central pillars (i.e., explore-exploit tradeoff and transfer mechanisms) of metacognitive learning. The four mechanisms constitute a tetrad of fundamental complementary processes, which coalesce to form the integrated bare-bones framework of metalearning, enabling young children's exceptional learning prowess. Figure 1 shows a topological representation of the formulated blueprint.

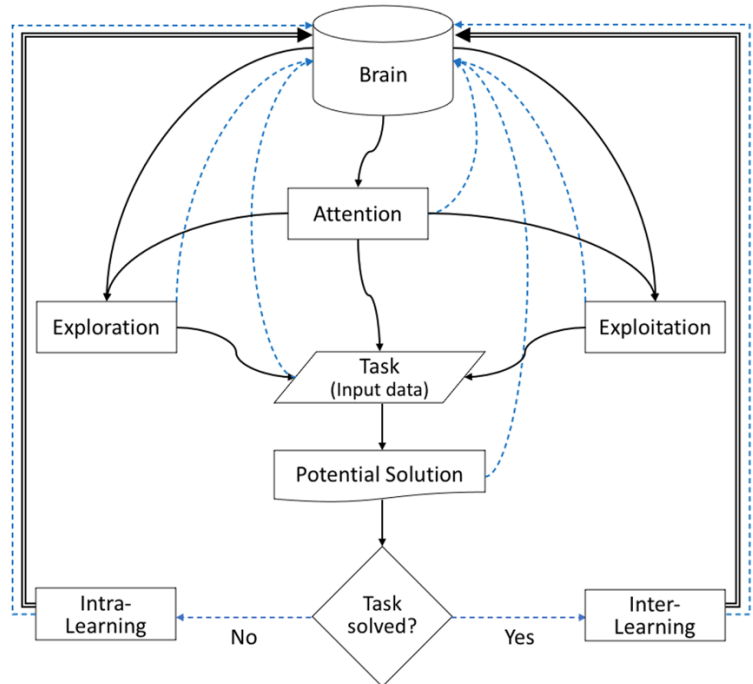

Figure 1: A topological schema of the proposed bare-bones metalearning framework with a process-centric configuration of the fundamental component mechanisms. The schema (herein, a single-task representation) depicts the metacognitive scaffolding as comprising four pillars or mechanisms—attention, explore-exploit tradeoff, feedback, and transfer—and emphasizes their centrality to the inspirational learning outcomes achieved by metalearning-enabled young children. Accordingly, the underlying mechanisms are displayed using a combination of rectangles as well as dashed and double-lined arrows outside the container that represents the developing brain in the schema. The explore-exploit tradeoff mechanism is represented by the exploration-exploitation pair of rectangles, whereas feedback and transfer processes are respectively delineated by dashed arrows and a grouping of intra-/inter-learning boxes and solid black double-lined arrows. The single-lined solid black arrows indicate process control flows and interactions between the elements illustrated in the schema.

As with the conceptualization of metacognition (Schraw et al., 2006), metacognitive learning or metalearning in the Figure 1 schema is theorized to emerge from the confluence of two factors: metacognitive knowledge (including the reflective awareness of morally grounded learning and the associated learning processes) and metacognitive regulation (including the management and control of morally grounded learning and the associated learning processes). The attention process, for the most part by itself and in collaboration with the other three fundamental enabling mechanisms, is the seat of metacognitive knowledge. Metacognitive regulation—which is known to encompass, in general, planning, monitoring, and evaluation (Schraw et al., 2006; Stanton et al., 2015)—is guided by the tetrad of core components of the proposed framework. In the developing brain, attentional awareness of a stimulus or learning task in a given problem context is likely primed by the autonomous goal directedness or motivation to resolve that task. The explore-exploit tradeoff and transfer mechanisms are respectively involved in learning and transfer strategy orchestration for implementing overall task learning within well-founded moral guardrails. The former process coordinates search and sampling patterns that represent a suitable compromise between exploration and exploitation, while the latter mechanism devises transfer paths for effecting the application of learning. The multimodal feedback process in the theorized framework gathers response signals through several feedback channels (*touchpoints*) to facilitate active monitoring (including deliberate examination of and manifest reflection on goal-directed strategies, strategy performance, and epistemic challenges). It works jointly with the transfer mechanism to foster overall learning improvements—by channeling feedback response profiles for intra-learning transfers—in the given problem context. Moreover, the attention mechanism operates in partnership with the feedback process to evaluate the multimodal response across the range of touchpoints within the framework and performs feedback priority scoring to prevent cognitive overload in the resource-constrained developing brain. Accordingly, once feedback is generated with the discovery of a potential solution, the attention process assesses the proposed solution and assigns a priority score contingent on the solution's suitability, efficacy, and moral viability in the problem context. Unless the potential solution solves the contextual task to the learner's satisfaction and is stored in long-term memory for application (via an inter-learning transfer) in a subsequent problem context, feedback priority informs strategy orchestration, adaptation, and implementation for the next learning iteration. In addition to evaluating potential solutions, the attention mechanism evaluates strategies associated with all four complementary processes of the metalearning representation for further process refinements. The continual learning improvements realized in the metalearning-supported developing brain are associated with higher levels of motivation, self-efficacy, learning efficiency, reflective awareness, and metacognitive regulation. These outcomes enabled by the four essential metacognitive learning processes potentially contribute to the surprisingly robust and morally sound cross-context learning accomplishments of young children. In sum, the theorized metalearning architecture in Figure 1 offers a process-to-outcome mapping—from the operation of the underlying complementary mechanisms to the impressive learning competence of the developing brain—as an explanation of young children's learning prowess. A potential strategy-driven pathway that engenders learning achievements representative of young children is outlined in Figure 2 under section 7.

    To learn about the world, the developing brain is tasked with harmonizing its exploratory and exploitative pursuits to acquire and process information from the environment that represents the world it is trying to comprehend. The act of exploration by the learners attempts to maximize their information gain from the environment while they remain cognizant of the nature and importance of available rewards. Conversely, the brain's preference for an exploitative strategy seeks to maximize the intrinsic value that the available rewards can yield through the information already gleaned from the environment (De Bruyn et al., 2020; Gualtieri & Finn, 2022). The contemporary constructivist learning view emphasizes the active involvement by the learners not only in their own learning activity (rather than them merely interacting with their individual cognitive processes), but also in displaying and applying acquired learning (including knowledge among other aspects of NK-B4P) in subsequent situations (Perkins & Salomon, 1992; Hajian, 2019). While children remain engaged in their individual learning processes, their propensity to carry learning from one context and utilize elements of that learning in another situation to influence its outcome (Perkins & Salomon, 1992), without diluting the fundamental nature of their acquired competence, is the focus and the essence of transfer. Although the learning transferred from the source context and that expected for the target problem may not be distinctly independent (Perkins & Salomon, 1992), the underlying relationship between the individual learning characteristics of the source-target context pair drives the notion of desired transfer[4]. It is this representation of transfer that is generally regarded as meaningful across academic disciplines, including education, machine learning (ML), neuroscience, and psychology. The affinity for desired transfer stems from what it engenders—the expansion of knowledge between source and target activities—even though impactful learning across contexts it unlikely without humans being actively vested in their own learning. Therefore, transfer, at least in the desired sense, is vital to the learning enterprise and requires appropriate cognitive and metacognitive strategy execution.

    The rest of the paper is organized as follows. Section 3 expounds the semantics of attention and its association with memory in forming a vital channel for fostering learning and the perception of learning in young children faced with constraints on their cognitive control. Section 4 discusses the significant aspects of their explore-exploit balancing act for morally centered learning and metalearning. Section 5 advances feedback as the metalearning mechanism that is critical for enabling the monitoring and evaluation of learning and the associated learning processes in the developing brain. Section 6 further expands on the notion of SRL to incorporate the process of learning transfer as the final pillar of the metalearning infrastructure for supporting knowledge expansion and abstract generalization in young children. Section 7 discusses the role of the integrated metalearning framework in the context of a feasible strategy-centric metalearning architecture with a strategy-to-outcome sequential mapping serving as a conduit for eliciting morally adapted learning realizations from the developing brain. Additionally, by considering the metacognitive underpinnings of the framework, this section distills the main implications for human brain-inspired neural networks that will predictably deliver

the next generation of responsible AI solutions. Finally, the section closes by examining opportunities for future metalearning research. Section 8 concludes the paper.

## 3. Attention mechanism

As a key learning enabler, the attention mechanism constitutes an essential pillar of the proposed metalearning framework and subsumes attention, memory, and a range of attention-mediated tasks. By guiding multiple sub-processes—including attentional focus, input feature extraction, chunking, data systematization, prior information filtration and regulation, memory clearing[5], perceptual specialization, and feedback assessment and prioritization—in the developing brain, the attention mechanism plays a crucial role in engendering reflective learning awareness in young children. In its raw manifestation, this mechanism represents an attention-mediated and moral-learning-centered adaptation of metacognitive knowledge: the classical building block of the metacognitive scaffolding underpinning the theorized framework. Attentional focus, the primary activity in the repertoire of sub-processes incorporated under this mechanism, is construed as an ordered process of cognitive reasoning (Hernandez and Amigo, 2021). Reasoning, in turn, is described as a cognitive activity which integrates attentional control with the assimilation of prior and prevailing knowledge in the brain to examine and corroborate ground truths (Hernandez and Amigo, 2021). The purpose of cognitive attentional control is to steer and regulate reasoning by concentrating cognitive resources on critical input fragments (Hernandez and Amigo, 2021) while filtering irrelevant details[6], to promote learning and improve learning outcomes (Shrager & Siegler, 1998). Research shows that young children tend to allocate their attention in a wide arc across the input signal, rather than discriminatingly attending to individual input fragments, which allows them to glean multiple features (both goal relevant and irrelevant) in the data (Gualtieri & Finn, 2022). Although the broad allocation of attention by young children seems counterintuitive, even countermanding their learning prowess, it may well symbolize a tactic to leverage cognitive control—a practice that regulates goal-driven behavior through attention and working memory management (Gualtieri & Finn, 2022)—and compensate for their cognitive limitations to extract superior learning outcomes. Moreover, research on attention-based categorical learning models that employ dynamic allocation of attention to facilitate object classification seems to support an initial wide-sweeping attentional focus followed by a gradual narrowing of attention (Niv et al., 2015).

Since children are known to have poor executive attentional control for directing attention (Gopnik, 2020) and reconciling inconsistencies among contending input signals (Garrigan et al, 2018), the attention mechanism supports activities that increase the exposure of these children to more complex (i.e., new or nuanced) representations with the aim of improving their attentional focus (Le et al., 2019). The mechanism allows attention, reflected across challenging cognitive tasks (including cognitive awareness, sensorially motivated response generation, strategy selection and implementation, and in-memory encoding and decoding), to direct reasoning by enabling selective focus on a subset of features of the input stream. Young children, therefore, confronted with cognitive constraints, tend to engage in sampling (Gopnik, 2020), i.e., chunking or disaggregation of information from the presented sample context, to work with smaller input subsamples instead of the entire input volume. Specifically, memory limitations direct children to parse and store the more tractable and informative input chunks that represent increasingly compelling and extreme associations among the features characterizing these samples (Gualtieri & Finn, 2022). In a way, then, the children's brain organically limits the dimensionality of information to relevant representations (although associated goal-irrelevant information may also be stored in memory, more likely than not if it is sensorially salient, conceivably to be used as a mnemonic trigger for easy recall of the corresponding relevant portion of the information). Niv et al. (2015) suggest that the brain's attentional control mechanism facilitates learning of multidimensional stimuli by directing attention selectively to a smaller subset of pertinent dimensions of a given problem context. With the data-efficient selection of relevant dimensions and the use of trial and error to update the contextual representation solely along these dimensions, this type of representation learning (Niv et al., 2015) may induce weakly organized and adaptable sparse semantic networks[7] in young children. A developing brain with an attentional control mechanism that tends to ignore weaker associations in data would likely support sparser, more flexible, semantic network topologies comprising loosely structured elements, each delineated by a vector of highly strong correlations among features extracted from the corresponding input chunks. The sparsity and adaptability inherent in the configuration of their knowledge structure, which affords attentional focus on the associations among objects, enables young children to avoid making false semantic connections between features that may appear to be semantically related (Gualtieri & Finn, 2022), thereby enhancing their cognitive reasoning outcomes.

Cognitive limitations, including constraints on cognitive control, influence the mental faculty to store and direct acoustic, visual, and other sense representations (allowing children to access their mental senses for creating and re-creating sensory inputs) as well as to retain and recall information over time. The type of information that is committed to memory and comprehended by humans is generally dependent (with the likely exception in the case of children in certain situations) on learning acquired from prior contexts (Gualtieri & Finn, 2022). Additionally, prior knowledge of fundamental ideas and actions is typically important to gaining new understanding of the more complex scenarios within the same context (Kang et al., 2019). Research, however, seems to indicate that the contribution of prior knowledge to new learning among children may be severely moderated and that lower levels of experienced learning proves advantageous to them for accurately identifying causal information (Gualtieri & Finn, 2022). Further, the brain's allocation of attention to information characterizing a given context impacts the memory of such information (Gualtieri & Finn, 2022). Working memory constraints influence the process by which children ascribe attention to information while inferring causal patterns in data (Gualtieri & Finn, 2022). By restricting the assimilation of earlier abstractions into current observations, these working memory limitations in children create ideal conditions for their brain to persist with empirical data to make causal inferences. Attributing importance to current observations rather than weighing prior knowledge in their inferential calculus yields unbiased causal inferences when the past and present information sources are inconsistent (Gualtieri & Finn, 2022). Other than ascribing differing weights across historical and current information, children may be inclined to give less (more) precedence to prior (existing) representations due to an underdeveloped neurocognitive apparatus, specifically, the declarative system (Gualtieri & Finn, 2022), which aids in their attention-based procedural or non-declarative learning. Long-term memory constraints in children may also contribute to the selective focus on relevant data representations. Although their less robust long-term memory inhibits new learning, the propensity to forget over time creates imperatives for supporting (precluding) more (less) pertinent information for learning (Gualtieri & Finn, 2022). Further, limited attention and memory enable children to utilize probabilistic data effectively for decision-making (Gualtieri & Finn, 2022). These constraints guide the attention-driven preservation of highly frequent data points in working memory, allowing children to advance the more recurrent option as the response. Tracking the most frequent response enables them to identify and pick the most probable outcome with a fair degree of regularity (Gualtieri & Finn, 2022). This capacity to discriminate highly likely patterns within the data, in the presence of cognitive restrictions, limits the children's ability to monitor and replicate low-probability features (Gualtieri & Finn, 2022), thereby aiding young children to develop memory for storing relevant information samples (Garrigan et al., 2018).

Some researchers postulate that prior knowledge tends to significantly influence memory recall (Gualtieri & Finn, 2022) and that working memory is tasked with processing new information before this information is passed along to long-term memory for storage (Du et al., 2022). Overloading working memory in young children degrades their task performance (Bosson et al., 2010), may impede their learning (Du et al., 2022), and even influence them to make uninhibited (including morally imprudent) decisions (Garrigan et al., 2018). Further, research indicates that despite applying attention broadly to maximize feature extraction, young children may find memory-intensive surveying of the entire input space and weighing of all data points in a given context challenging (Gualtieri & Finn, 2022). Although exposed to significant noise in data, they quickly adapt to consistent patterns in data via the attention mechanism that leverages their limited attention to facilitate perceptual specialization and rule learning. Whereas acutely tuned perception aids in recognizing and extracting input signals (Hadad et al., 2017), learning of rules (including heuristics) helps determine the structural regularity within the input space. Systematized knowledge tends to assist young children in sense-making of their environment, correlating new observations with prior information for certain tasks, making predictions, and drawing conclusions (Drigas et al., 2023). Resource-constrained cognitive control is believed to promote children's perseverative behavior, inducing greater predisposition for consistency in data (Gualtieri & Finn, 2022). The bias for input systematization introduced by enforcing order via their support for the more consistent forms in data accentuates the structural regularity of the input, allowing young children to tune in with their acute senses and align their representations of the data with the ground truth. For example, these children can correctly perceive causal patterns in the behavior of others. Specifically, young children can accurately switch between individual-specific and situation-specific causal attributions based on whether individuals behave consistently (even morally or otherwise) across tasks, thereby avoiding the *fundamental attribution error* (Gualtieri & Finn, 2022). Furthermore, their tendency to offer explanations consistent with empirical information generally remains immune to confounding context that may interfere with their sound reasoning and decision-making (Gualtieri & Finn, 2022). Reasoning, in general, blends the regulation of attention with the assimilation of knowledge to foster learning. In the present adaptation, however, attention comprises more than just a structured reasoning activity. Reasoning without moral centeredness or orientation can lead to suboptimal learning and antisocial decisions, conceivably making prevailing patterns of human behavior and social interactions unlikely (Dahl, 2019). The combination of rational and abstract reasoning, which facilitates problem-solving behavior, also determines reasoning that is driven by moral principles and understanding. Research suggests that both attention and working memory processes are critical for moral reasoning and development (Garrigan et al., 2018). Children as early as three months demonstrate moral proclivities, are guided by an understanding of moral rules, and negatively view decisions that entail harm (Garrigan et al., 2018) while achieving astounding feats of learning. Moreover, young children tend to decide morally, exhibiting prosocial behavior, when moral principles associated with a given situation are unambiguous and understandable. Nevertheless, moral decision-making becomes more challenging for them if their attention mechanism is unable to filter emotions accurately or when their working memory is overwhelmed (Garrigan et al., 2018).

To avoid cognitive overload and extract the most, given their cognitive limitations, young children being effective learners may be accessing latent cognitive processes or abilities that can be activated at their discretion to help mediate optimal learning. One such ability, integrated into the theorized metalearning framework, is operationalized via the putatively robust attention mechanism at work in young children. Besides allowing them to distribute their focus across the input space, the attention process in these children may also be triggered to function as a filtration mechanism (Garrigan et al., 2018) helping to discriminatively weed out prior knowledge that is incompatible with current observations. Additionally, a second latent cognitive process may be activated within the storage and recall system of the developing brain to alleviate the working memory footprint of attention-based tasks. This process is hypothesized to be associated with an intermediate memory cell—herein, designated as *transitional memory* and presumed to subsume Baddeley's (2000) episodic memory as well as ultra-short-term stores associated with sensory memory—which acts as a buffer between working and long-term storage. Accordingly, the transitional memory component within attentional control is envisioned as an off-ramp for offloading prior information from working memory to make space (for new information storage) and prevent interference. Once an information fragment is identified as a prior, it is buffered out to the transitional storage cell while new information is ingested and processed using the released working storage space. The prior information that conforms to current evidentiary patterns is presumed to revert to working memory when the developing brain is presented with specific stimuli (e.g., familiar voice prompts for language learning). In contrast, historical data that conflicts with current observations is purged from transitional memory. Research confirms that the human brain generally tends to allocate memory to new consistent information (i.e., details compatible with prior knowledge) rather than conflicting current observations (Gualtieri & Finn, 2022). Further, the flexibility to trigger the intermediate temporary storage may also allow young children to transfer highly correlated contextual information in working memory to long-term storage before the working capacity is overwhelmed. The attention mechanism is conceived to assist with the recall of such information in similar contexts by utilizing a memory-efficient information tracking and reactivation solution in the form of a metadata tag[8] consigned to working storage for each information relocation instance. We theorize that young children operationalize this approach as an essential alternative to using memory-intensive traces[9] of manipulations of mental representations in working memory. Under this conceptualization, information-specific memory is retrieved from long-term storage by engaging attention to instantiate the associated metadata tag rather than operate on the entire memory trace to reactivate it. Tags corresponding to unrecalled memories are transferred from working storage, while the associated informational content is relocated from long-term storage, into the transitional buffer to be purged. This attention-driven, *memory-clearing* process can potentially help young children negotiate cognitive constraints and limit cognitive load, an outcome that appears to be consistent with research in Gualtieri and Finn (2022) who document exceptional learning achievements of and memory-based advantages stemming from knowledge limitations in these children. By and large, the filtration function of attention and the constraints on memory together appear to limit knowledge acquisition by young children. Nevertheless, knowledge moderation plays to the advantage of these children as evidenced by their cross-domain cognitive accomplishments (Gualtieri & Finn, 2022) spanning sense-making, problem-solving, memory management, and moral decision-making. It should be noted that moral decisions of young children emerge within the proposed metalearning manifold as a corollary of their attention-based pursuits—aptly supported by the explore-exploit tradeoff, feedback, and transfer mechanisms—and not made independently of their engagement in cognitive and metacognitive activities.

The attention mechanism is indispensable for processing cognitive tasks, including information filtration and strategy execution. Attention may be directed, especially by sensorially salient stimuli, to shape perception without the benefit of experienced learning as part of a bottom-up information screening strategy (for instance, when young children encounter goal-irrelevant information). Alternatively, attentional capture can occur with a top-down filtering strategy that helps formulate interpretations of empirical information in concert with prior knowledge. The ability to regulate the orientation of attention on environmental stimuli is believed to be associated with working memory capacity (Fukuda & Vogel, 2009). Countermanding bottom-up attentional capture of visual stimuli can help working memory performance (Fukuda & Vogel, 2009), although curtailing top-down information filtering tends to engender innovative strategies in problem contexts (Gualtieri & Finn, 2022). As such, children display cognitive variation across diverse contexts and, consequently, engage attention in adaptively mapping their strategy choices to specific problems (Shrager & Siegler, 1998)—often applying overlapping maneuvers to address a specific problem (Rule et al., 2020; Shrager & Siegler, 1998). New strategies as well as existing approaches that are inadequately understood but remain to be executed are the subject of increased attentional resources, which improves their odds of being implemented flawlessly over a number of trials (Shrager & Siegler, 1998). The attention mechanism commits resources for strategy execution based on the relative importance (probabilistic assignments) of individual strategies to addressing specific problems in a given context. The assignment policy is designed to apportion probabilities in the order of ascending strategy utility, with the highest probability being assigned to the most useful approach, while the quantum of attentional resources committed to a strategy is constrained to vary inversely proportional to probability scores. The attention mechanism assigns lower (higher) probabilities to ineffective and inadequately learned (consistently useful and frequently implemented) approaches, thus allocating additional (fewer) attentional resources to approaches with lower (higher) probabilities until a designated frequency ceiling for strategy execution is reached or the corresponding problems are adequately resolved. If the upper limit imposed on the prevalence of a strategy is realized before the learner suitably[10] addresses a specific problem using the strategy in the given context, the mechanism calls for the synthesis and execution of a new strategy to attempt a resolution of the problem. This implies that children can synthesize novel approaches without earlier having employed ineffective or unsound strategies, which is congruent with research findings on strategy discovery in Shrager and Siegler (1998). Moreover, as the agent of strategy discovery and strategy application awareness, the attention mechanism is presumed to tag each strategy execution instance in working memory. But representations in working memory are susceptible to temporal decay (Towse et al., 2000), and catastrophic forgetting (CF)[11] can occur due to the interference among these representations. Once again, the attention mechanism is central to mitigating decay and CF as well as to fostering retention and recall of information (including knowledge of the strategies implemented) by operating on strategy execution metadata tags (in concert with the transitional store) to reactivate the corresponding mental representations in long-term storage through repetition or practice.

## 4. Explore-exploit tradeoff mechanism

When presented with sensory stimuli from the environment, young children deploy their attention to gain a wide-sweeping perspective of the associated input space. The aim is to optimize feature extraction while confining input dimensions to a range of pertinent features as they seek to subsample and standardize input samples, regulate the integration of priors, and select the corresponding perceptual information. However, human learning capacity essentially emerges from the balancing act between the explorative and the exploitative dimensions of the learner's search and sampling behavior (Gopnik, 2020). Accordingly, the support for NK-B4P[12] materializes through attention-directed activities subject to how young children probe the input sample space as well as formulate and test hypotheses across their designated regions of interest. The explorative attribute embodies both capacity and motivation for new learning, while the exploitative characteristic directs purposeful and premeditated action based on prior learning (Gopnik, 2020). Children are vastly explorative learners, actively searching for new information as well as persistently testing unconventional hypotheses and discovering novel solutions (Gualtieri & Finn, 2022), swapping initial explorative learning tendencies with subsequent exploitative demands (Gopnik, 2020). Information search and sampling leads to an inherent tension between exploration and exploitation: the explore-exploit dilemma (RID). The less-informed learner can launch an expanded (*high-temperature*) probe with limited evidence and continually test novel possibilities to acquire new knowledge. In contrast, a precocious learner may initiate a constricted (*low-temperature*) search and simply fine-tune the current hypothesis in tiny increments in the presence of fresh information (Gopnik, 2020).

In the explorative case, learners are likely to seek out high-level features that help explain their observations and discover new insights or possibilities within a given context, allowing them to make better abstract generalizations. However, explorative learners end up squandering resources by conceptualizing distant and improbable hypotheses if their initial premise is adequate (Gopnik, 2020). In the exploitative scenario, learners are likely to be more reliant on existing knowledge and make minor incremental updates locally to established premises to help craft an adequate solution. An exploitative strategy, nevertheless, can prove disadvantageous to learners if they fail to consider a distant hypothesis that could potentially engender a superior solution in the present context. One well-researched approach to resolving the RID is for learners to follow the *simulated-annealing* policy of leading with a high-temperature pursuit and converging to a low-temperature search (Gopnik, 2020). This general-to-specific search and sampling pattern enables learners to first explore the input space at a high level and sample from a broad range of possibilities that include divergent and unconventional hypotheses to maximize knowledge acquisition (Gopnik, 2020). Subsequently, it directs them to narrow their focus (Gopnik, 2020) on exploiting domain-relevant acquired knowledge to consolidate (Gualtieri & Finn, 2022) and exercise established options (Gopnik, 2020) for reward maximization. Research shows that young children effectively employ external exploration for active learning purposes: They not only engage in systematic exploration, pursuing and sampling new information about their environment, but also learn from their exploratory learning episodes (Gopnik, 2020). Curiosity is an important learner characteristic that drives exploration (Gopnik, 2020), a notion that is also supported by rational constructivism (Rule et al., 2020). Additionally, there is some support in literature for the idea that play enables children to perform internal exploration (including hypothesis search and sampling) as well as *competence exploration*, which encompasses problem-based strategy learning (Gopnik, 2020). Children adjust their principles and attitudes based on their observations from their explorative pursuits and use the underlying evidentiary patterns to generate fresh hypotheses. The explorative dimension of their search and sampling behavior, therefore, affords young children significant opportunities for learning as well as shaping what they learn.

Young learners (analogous to RL agents) may randomly adopt new behaviors or rules for efficient problem-solving (Rule et al., 2020). Utilizing heuristics to discover a solution through what seems to be a type of *random* exploration (Gopnik, 2020) is frequently disguised in children as a form of *directed* exploration: a distinctively crafted, goal-driven activity (Rule et al., 2020). Young children may be motivated to learn moral rules concomitantly with their directed exploratory pursuits for two reasons. First, moral rule learning may help to regulate behavior as children attempt to align theirs with existing prosocial norms. Research reveals that young children display moral inclinations consistent with principled behavior while exploring their environment, learning and following rules prescribed by grown-ups in problem contexts that require moral reasoning and decision-making (Garrigan et al., 2018). Moreover, young children are sensitive to moral repercussions while evaluating their explorative actions and even go so far as to perform a cost-benefit analysis of harm, although they invariably cast explicitly beneficial actions associated with detrimental outcomes in a negative light (Garrigan et al., 2018). In essence, learning moral rules can lend consistency in recognizing moral imperatives in diverse contexts, guiding their morally oriented actions and controlling their behaviors to make them conformant. Second, learning and employing heuristics centered around moral decisions may serve to ameliorate the RID as the developing brain adaptively trades off controlled exploitation for unconstrained exploration, effectively driving moral rule learning. Changes in neural plasticity (i.e., responsiveness of the brain to external input), which allow the human brain to switch between excitatory activity supporting synaptic network reorganization and inhibitory reflex reaction fostering stabilization of neuronal connections, can be immensely relevant to effectuate this trade-off. In periods of plasticity and perceptual expansion during cognitive development, the human brain is acutely receptive to external input (Gualtieri & Finn, 2022). Limited cognitive control in young children additionally attenuates inhibition and intensifies explorative search and sampling behavior, driving these children to consider alternative solutions that are more divergent and less substantiated (Gualtieri & Finn, 2022) but morally compliant (Garrigan et al., 2018). An extended period of exploration can also engender an entire range of cognitive skills (Gopnik, 2020), including moral reasoning.

Furthermore, since young children are generally free from the notion of functional rigidity associated with the general perception of common objects, their explorative problem-solving approach involves attributing divergent or unconventional functional applications to these objects allowing the children to realize innovative solutions that can be highly effective at times (Gualtieri & Finn, 2022). Another piece of their cognitive milieu that supports this dynamic is their rapidly developing procedural neurocognitive learning system which competes with and is aided by its immature and slowly developing declarative counterpart. An underdeveloped declarative system maturing deliberately over time, counterbalanced by the swiftly maturing procedural system, decreases (increases) the likelihood of children's reliance on prior (existing) functional information to discover novel solutions to problems prone to the functional-rigidity bias (Gualtieri & Finn, 2022). Reduced dependence on prior knowledge allows greater scope for exploring novel hypotheses (Gualtieri & Finn, 2022), including those underlying moral centeredness and related heuristics, which can help discover morally responsible solutions. When current experience generates a viable solution, the plasticity of the brain enables the introduction of new (Gopnik, 2020), as well as the reorganization of existing, synaptic connections to reinforce experience-based learning (Gualtieri & Finn, 2022). Accordingly, children modify their behaviors to act morally while exploring in a highly variable environment characterized by a wide range of complex options. If the situation is vastly predictable and can be represented by a limited set of conventional possibilities, young children may be equipped to adjust their explore-exploit patterns and even engage in learned rule reversals (Gopnik, 2020) for the current context. In this situation, the brain will act to reinforce certain neural pathways while pruning redundant connections yielding a more controlled and efficient, yet rigid exploitative state that curtails salient network reorganization (Gopnik, 2020). The resultant exploitative behavior may help to ensure that young children focus their attention (supported by the feedback mechanism within the proposed metalearning framework) on search and sampling behaviors that maximize known rewards. The exploitative adjustments made by children may also allow them to optimize their moral alignment assessments while performing goal-directed tasks, especially if the given reward structure incorporates incentives for moral compliance. Although perceptual contraction can set in rapidly during exploitative periods accompanied by the stabilization of synaptic networks, children can leverage their experiences to reorganize and revise acquired learning (while operating within **moral** guardrails) even well beyond the designated span of neural plasticity accompanying cognitive development (Gualtieri & Finn, 2022). Consequently, with neuronal plasticity and adaptability being integral to resolving RID as part of the theorized metalearning framework, the developing brain may be capable of generating explore-exploit tweaks in real time to a certain degree by making corresponding adjustments in its synaptic networks in crucial problem contexts.

## 5. Feedback mechanism

Learning, including morally oriented knowledge acquisition, for contextual problem-solving by young children materializes from the resolution of RID via context-dependent adjustments in their explorative and exploitative behavior. The explore-exploit tradeoff mechanism in young children zeroes in on the search and sampling approach that extends and optimizes attention-based probing and subsampling of the input sample space. Therefore, the harmonization of exploration and exploitation works in concert with the attention mechanism to abstract away the complexity in data and distill the salient knowledge kernel which accurately represents the underlying patterns in a given input sample. The neuroplasticity of the human brain aids and abets this tradeoff mechanism. When faced with unpredictability, young children may be able to trigger periods of plasticity—so that their malleable brain remains responsive to stimuli and supports neural network growth and restructuring—which allows them to adapt their search and sampling behavior in accordance with observed patterns in data. Accordingly, children engage in exploration to maximize new learning in the given context.

Conversely, in more routine contexts, young children's experience-driven search and sampling adaptation induces pruning as well as reinforcement of neural connections, curtailing brain plasticity and preventing network restructuring to yield a more inflexible yet stable synaptic manifold. By acting exploitatively as part of a neuroplastic response to their environment, these children then focus on optimizing rewards anticipated within the presented context while conducting themselves in a manner consistent with established moral principles. While young children make adaptive choices regarding their explorative and exploitative strategies to learn a context-relevant hypothesis space[13], they refine their problem-solving skills through critical examination of and explicit reflection on the acquired learning and the associated learning activities. Consequently, a feedback apparatus is fundamental to processes and strategies that involve self-reflection (Drigas et al., 2023) and moral decision-making. For example, when typically developing young children encounter a choice between playing with a light-up ball (specifically designed to deliver an exciting sensory experience) and helping a sick adult reach for a cup of water (generally perceived as a moral imperative), a performant feedback mechanism is likely to elicit self-monitoring and attention-guided reflective awareness, which directs them to pick the morally correct choice. Within the perspective of learning to learn morally grounded solutions to problems in diverse scenarios, the feedback system is responsible for the orchestration of monitoring operations that enable the monitoring of learning as well as the attendant processes and strategies. The role allows the system to generate adaptive response profiles that constitute feedback circuits of insightful data, including information on state representations, stimuli-driven behavioral patterns, introspective learning strategy assessment, and other internal response dynamics, extracted during morally aligned learning enterprises. The developing brain in the proposed metalearning interpretation is theorized to perceive these dynamic profiles as supplemental inputs in a feedback loop utilized for tracking progress, adjusting strategies, and regulating learning. The process is conceived to fine-tune learning in young children by harmonizing internal insights and external perspectives, engendering performance improvements while helping to mitigate CF during purposeful undertakings aimed at resolving the designated task without compromising moral priorities.

Given the spate of technology innovation over the last two decades, some research has been directed at investigating and exploring the use of technology frameworks and technology-assisted interventions, which include feedback provisioning, to foster self-refection in children. Technology deployments to support learning in schools potentially create greater metalearning activity which, in turn, promotes superior learning outcomes that are reflected in the application of feedback-enriched solution iterations across problem contexts. Drigas et al. (2023), in examining evidence-based research on the use of metalearning strategies with smart automation, find that learner-driven feedback facilitated by technology is crucial for supporting self-reflection and adaptive learning. Deng et al. (2011) disclose that students who actively participate in learning via computer-assisted technology mediation with instantaneous feedback are likely to succeed in advancing their conceptual understanding and metacognitive abilities spanning metacognitive knowledge and regulation. Interactive educational software paired with AI-based learning technologies are being employed to provide real-time feedback to help students monitor their progress and prompt them to reflect on their learning (Chen et al., 2020; Drigas et al., 2023). The results of a meta-analysis study by Van der Kleij et al. (2015) reveal that the more explanation-oriented the feedback in computer-based learning environments is, the better the primary and secondary school student performance on learning assessments tends to be, especially, in higher-order (i.e., metacognitive) and transfer learning contexts. The feedback apparatus, leveraged in the developing brain to help monitor learning and the related cognitive activities with the aim of promoting self-reflection, is conceived to operate collaboratively with the attention mechanism in parsing and coordinating both internal response patterns and external feedback.

The importance of feedback in SRL studies (Deng et al., 2011; Drigas et al., 2023; Van der Kleij et al., 2015) establishes the criticality of the feedback mechanism in the posited metalearning framework. The information that learners receive about their state of learning and performance in various contexts relative to their desired learning outcomes and designated contextual objectives is critical to the refocusing and realignment of learner efforts for goal realization. Feedback about directing attention for subsampling the input space, the degree of systematization needed to facilitate the selection of more adaptable and consequential input chunks, balancing exploitation and exploration to learn potential solutions, the efficacy of new strategies and potential solutions, and the individual results of applying the learned solutions to specific problem contexts constitutes an integral component of learning to learn with moral integrity. Accordingly, the feedback mechanism within the posited metacognitive scaffolding is conceptualized to receive feedback signals from multiple *touchpoints* across the neurobiological landscape and communicate these signals to the attention mechanism that ascribes relative importance scores (i.e., numerical weights) to the individual instances of the incoming feedback array. The higher the weight associated with a feedback instance, the greater is the assigned likelihood of it being incorporated back into the metalearning manifold to hone learning. Salient learning feedback data are assimilated as ancillary input signals for informing both real-time reactions and deliberate strategy improvements (via critical reflection and perspective taking) for facilitating regulation of learning and engendering highly effective learning outcomes over time. Moreover, the capacity to reflect on learning progress and strategy utilization as well as to consider perspectives of responsible adults in problem-solving pursuits supports metacognition, as children are motivated through self-reflection and perspective-taking to assess the impact of their behaviors on others. Further, the feedback mechanism may foster the development of emotional and moral intelligence by equipping young children with the metacognitive skills, empowered by self-reflection, to understand the emotional and moral consequences of their actions. The same intelligence promotes the regulation of emotions and moral learning in these children.

## 6. Transfer mechanism

Young children are capable of incredible acts of learning over a wide range of contexts. In this paper, we propose a novel bare-bones metalearning framework that offers a theoretical perspective on how these children might achieve impressive learning outcomes and formulate effective solutions to problems without ignoring moral priorities. Their attentional control and explore-exploit adaptations—representing two essential elements of the metalearning architecture—optimize search and sampling of the input space, reducing data dimensionality and cognitive load to generate novel solutions in a given problem context. The feedback mechanism, another mainstay of this framework, enhances the children's problem-solving performance through recursive trials which entail feeding back outputs as inputs and concatenating output-input-output instances to form learning circuits. These feedback loops may be utilized within the framework for engendering a morally aligned solution that satisfactorily addresses the problem in the presented context. More than discovering a viable solution in siloed contexts, however, learning comprises the ability to demonstrate learning transfer (Perkins & Salomon, 1992) beyond knowledge acquisition and solution discovery. The human capacity for generalization and adaptive decision-making, enabling the formulation of predictions in a target context facilitated by the transfer of learning from a source context, is central to inducing successful learning outcomes (Niv et al., 2015). Accordingly, a transfer mechanism to support cross-context and meaningful learning in young children constitutes the final pillar of the proposed metalearning infrastructure.

An effective transfer mechanism allows meaningful learning transfers, that is, transfers that are additive. When knowledge acquired from a source context enhances the learning outcome of a target context such that the source-to-target learning transfer grows the learning output synergistically, the learning quantum accumulated from sequentially addressing (with transfer) the source-target problem combination is greater than the sum of the individual learning quanta that would have been achieved from an independent resolution (without transfer) of the problem pair. The transfer mechanism, as an integral component of the theorized metalearning representation, enables young children to evoke additive learning transfers via *positive* transfer episodes. A positive (negative) transfer occurs when the performance or learning outcome in a certain situation gets better (worse) due to learning in a prior context (Chen & Daehler, 1989; Perkins & Salomon, 1992). Clearly, cross-context learning is adversely impacted (producing ineffective learning outcomes) by negative transfers. In fact, Chen and Daehler (1989) observe negative transfers for problem-solving tasks undertaken by 6-year-olds faced with divergent solutions to source and target problems, particularly, in learning contexts that entail a *far* as opposed to a *near* source-to-target transfer[14]. According to Perkins and Salomon (1992), negative transfers generally present a challenge during the initial phases of learning a new target activity, but subsequent experience-driven learning can help mitigate such adverse transfers. The transfer mechanism may assist with moderating the inability of young children to transfer learning across, especially, disparate contexts, helping to accelerate learning from experiences, by affording children opportunities for exploring specific actions[15] or strategies. Negative-transfer-mitigating strategies may include developing immersive experiences in activities associated with the source problem, identifying relationships across isomorphic problem contexts, abstracting fundamental characteristics from a given problem context, becoming active participants in the monitoring and reflection of personal cognitive processes, drawing metaphorical or analogical correspondence over problem domains, and cultivating acute awareness of one's environment and the problem-solving approaches that one adopts (Perkins & Salomon, 1992).

Successful cross-context (i.e., positive) learning transfers are the desired transfers that enable learners to expand their repertoire of both knowledge and experience. These transfers could play a pivotal role in helping young children develop their learning capacity, thereby allowing them to generalize across contexts. As proposed in Zhang and Zhang (2019) and examined to some extent in Perkins and Salomon (1992), self-regulation and self-monitoring in the SRL context can drive learning capacity expansion. Taken together, the co-primacy of learning regulation and learning transfer may help support the selection and adaptation of cognitive and metacognitive strategies by children in diverse problem-solving settings to reinforce meaningful learning outcomes. Perkins and Salomon (1992) propose two channels (referred to as the *classical transfer paths* in this paper) for driving learning transfers. Their low path transfer[16] is characterized by relatively automatic (*reflexive*) response regimes or practiced strategies that are easily activated in likely near-transfer scenarios (i.e., where source and target contexts have an adequately high number of common features). When practiced routines fail to get triggered automatically, the brain may resort to heuristics (i.e., mental representations) or improvisations for driving target outcomes. In contrast, the high path transfer is generally a non-reflexive response requiring awareness-directed abstraction of cross-domain principles and purposeful exploration of likely associations across potentially dissimilar contexts. Aside from the two classical transfer paths, wherein transfers are either reflex-driven or deliberate, the theorized metalearning framework supports the emergence of two novel transfer channels (*developmental paths*) that young learners are likely to incorporate in certain contexts. Integrating the developmental modalities of *free-space* and *well-trodden* path transfers with the classical transfer channels yields a more inclusive transfer of learning theory. A broader theoretical scope may help uncover additional pathways and illuminate finer details about the conditions under which young children effectuate transfers along a range of paths in a variety of learning contexts. Free-space path transfer, for instance, is expected to serve near transfer events, especially, when learning acquired in the source context is likely to attenuate with time due to any number of mitigating factors (including cognitive limitations, learning disabilities, transfer delays, and information interference). This attenuation occurs as learning progresses through *free space* (i.e., contexts) for potential application in the target context, with the learning quantum and quality degrading along the way in the presence of mitigating factors. Research shows that young children often experience forgetting, a form of information attenuation, because of long-term memory limitations (Gualtieri & Finn, 2022) and that their performance tends to degrade as their memory is overworked with problem-solving tasks (Bosson et al., 2010). In free-space path transfers, the learning signal across similar contexts can be boosted through appropriate interventions.

Well-trodden path transfers, in contrast, may be employed by young children in near and far transfer situations (although more likely in the case of the former) when learning can be applied, without much practice or deliberate abstraction, in familiar as well as novel contexts after observing adults demonstrate cross-context application of learning. Research shows that young children are equipped with social-cognitive skills for interacting with and imitating familiar actions as well as learning new feats from others (Callaghan et al, 2011). Further, these children are more inclined to learn selectively by mimicking adults rather than their peers (Rakoczy et al., 2010). Therefore, well-trodden path transfers may be ideal for settings where imitation learning is common, indicating potential applications within the social and community learning landscape. A preliminary conscious effort at imitation in a problem context can become so predictable or ingrained in behavior through repetition as to cause the learner to abandon a well-trodden transfer strategy and instead transition to a reflexive response (low path transfer) scenario in a subsequent near or far context. Additionally, a deliberate effort to obtain a symbolic structural representation of the source context for application in a target context may not accomplish a classic high path transfer due to ingrained mitigating factors (such as individual learner characteristics and source-to-target temporal separation) or evolving moderating influences (including procrastination and dwindling motivation). In this situation, the learning agent may be inclined to abort the high path transfer and convert to a more suitable transfer option. A transfer path shift—termination of the original transfer channel and (or) conversion (of the intended path transfer) to, for example, a free-space path transfer—effectuates an attenuated version of the originally learned representation to be transferred. In the absence of appropriate interventions, this transfer instance may lead to suboptimal performance in the target context. To be consistent with the morally guided learning efficacy evidenced in the outcomes achieved by young children, therefore, the metalearning architecture is theorized to facilitate learning transfer adaptability, enabling transfer path shifts between pairs of the classical and the developmental transfer modalities in response to contextual and learner characteristics. Any conceptualization that calls for the immutability of learning transfer channels is as counterintuitive as the notion that proximity is precisely codified as *near* and *far* in transfer scenarios, the latter idea being dismissed by Perkins and Salomon (1992). Accordingly, the proposed metalearning framework affords young children opportunities to execute transfer path adaptations and actively leverage intra-learning transfers that serve to advance learning iteratively through learner-motivated continual practice (facilitated by the feedback mechanism), or deliberate explorative pursuits (supported by the explore-exploit tradeoff mechanism), or both. The accumulation of learning, which occurs as a consequence of a positive transfer toward resolving the known problem context, is assigned to the metalearning-enabled brain's long-term memory via an inter-learning transfer for future deployment in unknown contexts. Without transfer, knowledge acquisition may not be realized, and a cascading failure of learning outcomes throughout the problem space is likely—paving the way for a catastrophic learning crisis—in the absence of knowledge. Therefore, within the posited metalearning infrastructure, the transfer mechanism plays a pivotal role in ensuring a balance between the primacy of knowledge and the preponderance of process.

## 7. Discussion

Young children are exceptional learners, effectively crafting hypotheses for and solutions to the problems they encounter in their environment even before exposure to formal instruction in school (Gopnik, 2022). The capacity to apply their knowledge and skills both reflexively and deliberately in diverse circumstances, despite their cognitive limitations, equips these children to engender successful autonomous and morally grounded learning outcomes across multiple domains, including acoustic, linguistic, and visual. This raises two important questions, one related to the enablers of young children's astonishing cross-context learning ability and the other to the prospect of human brain-inspired artificial intelligence (AI) converging to morally guided understanding representative of young children. The first question on what might empower young children to learn surprisingly well, helping them to deploy what they learn across problem contexts, is at the heart of researching and uncovering the fundamental enablers of the children's ability to learn about their morally oriented learning. The second question focuses on how addressing the first may help guide morally grounded innovations in human brain-like synthetic intelligence. We address the first question by devising a novel integrated metalearning infrastructure, which captures and explicates the functions of and the interactions among four underlying mechanisms that enable learning while generating learning awareness in the developing brain. The proposed metacognitive scaffolding embodies a self-regulated learning (SRL) representation comprising task learning, learning about task learning, and self-motivated task understanding. An SRL architecture would be deficient as an explicatory framework so long as it occludes or discounts the individual contributions—including attention-memory interaction, explorative-exploitative plasticity, recursive feedback, and transfer learning—of an irreducible tetrad of enabling mechanisms to goal-driven learning in the presence of epistemic limitations. Accordingly, in this paper, attentional control, exploration-exploitation search and sampling dynamic, feedback-guided regulation, and intra-/inter-learning transfer mechanisms are interlaced into a bare-bones metalearning framework. Essentially, a pair of learning enablers based on individual adaptations of two indispensable classical dimensions (awareness and regulation) of the metacognitive scaffolding are integrated with two other crucial enabling mechanisms (explore-exploit tradeoff and transfer) to conceptualize a novel representation of metalearning. Based on available information, this paper is the first to integrate four fundamental complementary mechanisms into a unique parsimonious model of metalearning which can be utilized to illuminate the underlying processes responsible for endowing young children with the competence to perform amazing learning feats across contexts. The proposed framework can be visualized using a topological process-oriented schema (Figure 1) as well as a multilayered strategy-centric representation (Figure 2). Using the latter schema, we summarize a plausible configuration of strategy layers (i.e., a streamlined sequence of strategy implementation steps) which induces metalearning-enabled, high-performance learning in young children.

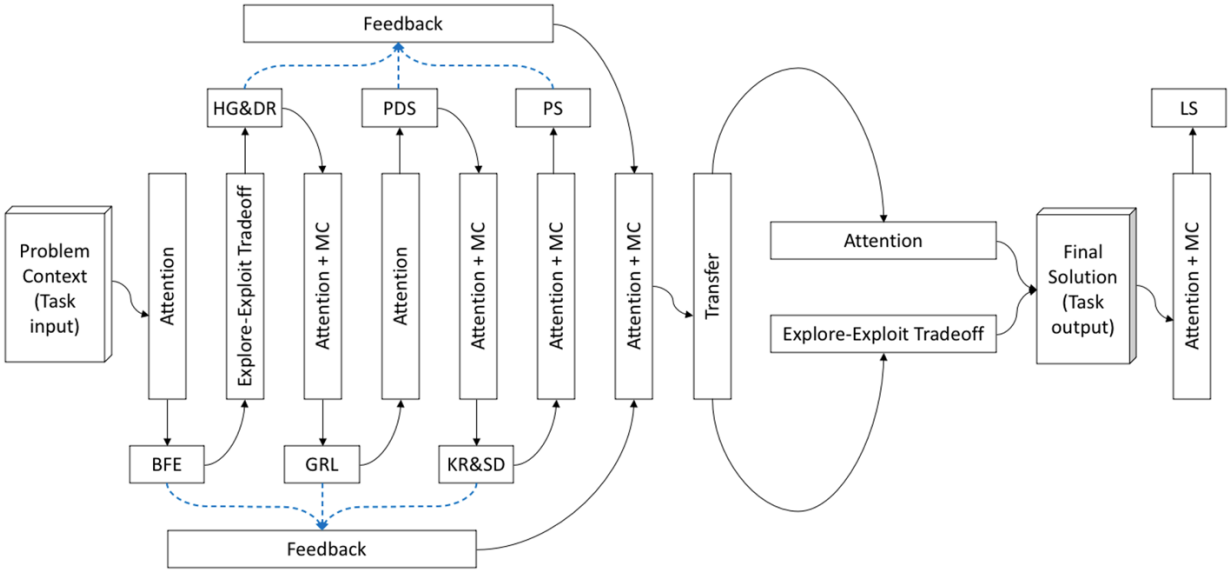

Figure 2: A multilayered representation of the single-task metalearning schema (Figure 1). The schema herein depicts a strategy-centric, layered configuration that emerges from mapping specific strategies executed in the developing brain to their corresponding outcomes achieved toward solving the task in the given context. The solid black arrows represent data flows from the given learning problem (task input) to the attention-driven strategy layer executed on the final solution (task output) to yield the learned solution, while the dashed arrows delineate flows underlying strategy monitoring from multiple strategy touchpoints (discrete strategy-outcome occurrences that provide opportunities for metacognitive awareness, perspective taking, and strategy discovery within the metalearning framework) to the feedback mechanism. Starting from left to right in the schema is a potential pathway for enabling metalearning coupled with goal-directed behavior and task learning. At the outset of this path, the problem context is subjected to the attention strategy layer (or, conversely, an attention strategy is triggered to act on stimuli within the problem context) to shape learner perception while learners extract a wide range of features across the input space. This attention strategy produces an outcome symbolized as BFE (*broad-spectrum feature extraction*). Moving to the right in the multilayered schema reveals additional outcomes, such as HG&DR (*hypothesis generation and dimensionality reduction*), GRL (*goal-relevant learning*), PDS (*pertinent data systematization*), KR&SD (*knowledge regulation and strategy discovery*), PS (*potential solution*), and LS (*learned solution*). MC denotes *memory clearing*, a strategy for managing representations in memory using the postulated transitional store, which is combined with attention layers in the schema.

*7.1 Strategy-centric metalearning representation*

Within the strategy-centric metalearning architecture (Figure 2), young children (especially, typically developing learners) are considered to be autonomously driven to focus their attention strategy on the learning task in a given problem context, making a broad sweep of the input space to glean as many contextual features as possible despite their cognitive restrictions. They subsequently execute a search and sampling pattern, directed by their context-relevant strategy that elicits a suitable tradeoff between their explorative and exploitative pursuits, to reduce the dimensionality of multidimensional data across the input space and learn a relevant hypothesis set. The input data are effectively scaled down to primarily goal-relevant representations, while their attention-enabled strategy combination of information filtration and memory clearing commits the residual goal-irrelevant information to transitional memory and consigns the corresponding resource-efficient metadata tags to working storage for on-demand retrieval or designated purge. The next strategy layer in the framework involves engaging the attention mechanism to leverage the limited attentional resources in the developing brain to induce perceptual specialization and rule learning for regularizing the pertinent representations of the dimensionally reduced input space while fostering awareness of moral imperatives. Thereafter, attention with memory-clearing strategy use directs attention-driven information filtration and prior knowledge regulation to identify priors inconsistent with current empirical data for removal in a memory-clearing exercise. As young children deconstruct, systematize, and regulate input data, they actively and recursively assess the informational content and moral soundness of relevant data fragments, thereby learning about their morally centered process and strategy outcomes via the feedback mechanism. These children may encounter technology aids to assist their developing brain with unraveling constructive insights, tracking learning progress, and promoting self-reflection, enhancing their overall learning experience. For each feedback instance, a strategy integrating attention with memory clearing is triggered to perform feedback priority scoring to identify salient feedback profiles and prevent cognitive overload. The higher-scoring feedback instances are recursively driven to the transfer mechanism which helps to loop them back by implementing appropriate transfer strategies. These include intra-learning transfers to both attention and explore-exploit tradeoff mechanisms to effect adaptive modifications to their respective strategies. The strategy combination of attention and memory clearing is re-engaged in the next layer to distil an appropriate solution representation from the goal-relevant, experienced-regulated hypothesis space. A potential solution emerges as competing solution encodings are rejected and relegated for clearing. The posited framework relies on the neuroplasticity associated with the developing brain to enable continual strategy adaptations for learning a morally grounded and decisive solution that suitably addresses the problem context. Finally, the developing brain once again synthesizes and deploys an attention strategy with memory clearing to label (using an appropriate metadata tag in working storage) the memory trace of the acquired learning representation (i.e., learned solution) in the existing context. While metadata tagging is instantiated, the related actual memory trace is pushed into long-term storage for subsequent application (i.e., via positive inter-learning transfers) in other problem contexts. The triumvirate of motivation, task learning, and metalearning manifests itself through the execution of context-driven strategies related to the fundamental enabling mechanisms within the conceptualized metacognitive scaffolding, acutely priming the metalearning-enabled brain of young children for exceptional performance across a plurality of subsequent near and far learning tasks and contexts.

*7.2 Process-driven multitask extension of the metalearning representation and AI implications*

The postulated framework views young children as self-motivated neurobiological agents performing key learning functions. Specifically, the developing brain dynamically deploys attention, adaptively balances unconstrained exploration against measured exploitation to fine-tune learning strategies, actively generates feedback from monitoring and evaluating these strategies, and engages in learning transfers in response to feedback. Understanding how the developing brain is enabled to achieve the triadic motivation-learning-metalearning configuration by executing an appropriate sequence of strategies, can offer a roadmap for designing efficient and responsible neuromorphic AI systems. The SRL triad, evident via strategy implementations associated with the fundamental processes of the proposed metalearning framework, comprises the prerequisites for empowering the host AI system to deliver morally compliant performance across learning contexts. Accordingly, the sequence of strategy execution stages can be conceived as applications of functional (i.e., specifically tasked attention, explore-exploit-tradeoff, feedback, and transfer) layers in multilayered neural networks that are at the heart of AI solution engineering. Alternatively, the functional applications can be aggregated under each core mechanism to yield a process-centric architecture. Figure 3 extends the single-task metalearning schema presented in Figure 1 to a multitask, single-context representation in the AI realm.

A metalearning framework that can serve as a lens to examine and expound young children's epistemic competence for morally inclined learning in diverse circumstances—including acoustic, linguistic, and visual contexts—has important implications for responsible AI development. In particular, the paper draws on the correspondence between the developing brain and its brain-inspired AI counterpart in the SRL context to infer four main associations for guiding advancements toward morally conformant AI.

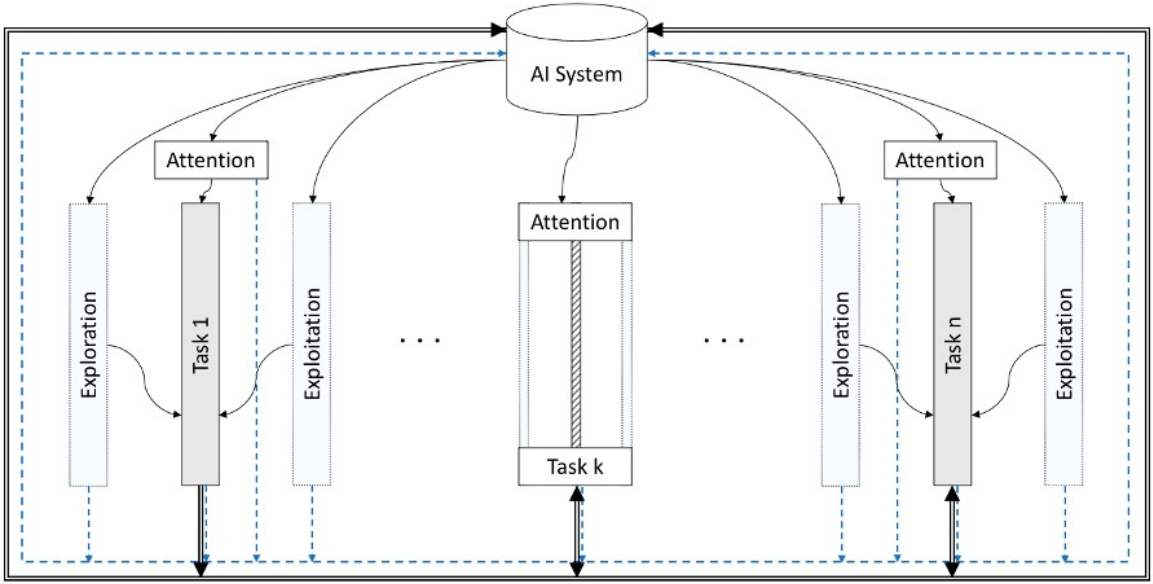

Figure 3: A multitask, single-context metalearning schema which is an adaptation of the process-driven, single-task (single-context) metalearning representation. The schema in this figure comprises a nesting of n-single-task metalearning schemas, with the associated learning tasks (within the same problem context) being respectively designated as *Task 1* through *Task n*. The task related to the intervening metalearning block displayed in the center of the schema is labeled arbitrarily as *Task k*. The solid black arrows represent process control flows and inter-element interactions, while the dashed arrows denote feedback channels that constitute a shared feedback mechanism. The solid black double-lined arrows, denoting learning transfers associated with each block, connect to a common transfer mechanism. While a dedicated AI system (including the hardware to support and accelerate neuromorphic computing) shares both the feedback and the transfer mechanisms across a range of tasks, a separate pair of attention and explore-exploit tradeoff processes is assigned to each block for preserving the schema's functional validity (each task may require specialized attentional focus and individualized explorative-exploitative dynamic). Working with the assumption that *Task 1* is the very first task that the neuromorphic AI system is seeking to resolve, the initial metalearning block can only generate, but not receive, an inter-learning transfer. Hence, the associated solid black double-lined arrow is represented as a single-headed arrow. Once *Task 1* is learned, the associated positive transfer of learning from the current block is accomplished by the transfer mechanism via an inter-learning transfer to the next block responsible for learning *Task 2*. Except for the initial block, all others are assumed to be both contributors and recipients of learning transfers. Therefore, block learning transfers beyond *Task 1* (i.e., transfers to *Task k* and *Task n* in this figure) are shown with a double-headed arrow. Although there is an arbitrary cap of n tasks, n being some threshold imposed by the system's resource limitations, the brain-inspired architecture can facilitate the transfer of the accumulated task learning at *Task n* to potentially any task in a subsequent context (not mapped in the schema) in the absence of mitigating factors.

*7.2.1 Resource constraints in AI-embedded solutions that integrate the posited metalearning architecture can be exploited to gain certain advantages*

As with the emergence of children's epistemic competence in the presence of and facilitated by cognitive limitations, resource-constrained AI systems based on brain-like neural network architectures that incorporate the proposed metalearning framework can generate both efficient and effective, morally adapted learning outcomes. The streamlined architecture of the suggested framework underscores the value of resource limitations, essentially endorsing the perception that cognitive constraints may not necessarily be a disadvantage to learning. In fact, the presence of cognitive limitations in young children may act as requisite motivation for the developing brain to deliberately execute and selectively adapt its attention, explore-exploit tradeoff, feedback, and learning transfer strategies with the purpose of optimizing learning outcomes without having to overlook or trivialize moral imperatives in the given problem context. For human brain-like computing incorporated in deep learning (DL) applications, especially the AI-based embedded systems that are ubiquitously deployed in the Internet of Things (IoT) technology space (Huang, 2022; Mehmood et al., 2022; Mehonic & Kenyon, 2022), future innovation may be better served if it is modeled after the metalearning experiences in young children. By integrating the proposed metalearning architecture into existing and prospective DL applications, the framework's inherent strengths can translate into operational efficiencies for the host or embedding AI technologies. Resource-constrained computing problems, in particular, can be mitigated for future edge-AI innovations that incorporate the metalearning scaffolding conceptualized in this paper. Emulating the fine-grained functions of the four complementary mechanisms of the theorized metalearning framework in metalearning-augmented IoT and other resource-limited AI solutions can power a context-contingent transformational shift in operational focus and resource utilization. The metalearning-based augmentation of the adapted smart solutions can employ the explore-exploit tradeoff mechanism to effectuate adjustments in the functional orientation—from just accuracy (i.e., model prediction and ground truth correspondence) or efficiency (i.e., computational speed and economy) to an appropriate compromise between accuracy and efficiency—of the DL models driving these technologies. While modifications in the desired model outcome of an augmented AI solution are enabled in specific contexts, the solution's attention component can deploy a caching routine that leverages smart tagging to retrieve frequently accessed data, reducing latency and recurrent processing. Metadata labeling and transitional cache-based memory clearing can also alleviate computational loads by lowering the memory overhead in resource-deprived AI systems. Additionally, the AI attention mechanism can implement an intelligent resource allocation and scaling policy that learns and facilitates dynamic resource distribution with adaptive resource pool adjustments to improve efficiency without compromising accuracy. Aided by a rapid multimodal feedback component and a robust transfer mechanism, metalearning-enabled AI solutions can further benefit from persistent resource-aware and morally perceptive learning. As human computational demands for processing greater volumes of unstructured data in real time grow, brain-inspired AI architectures supporting the rapid interplay among integrated metalearning components to help accelerate efficient computing (while enabling accurate, morally adapted outcomes) in resource-constrained settings could become the norm.

*7.2.2 Plasticity and contextual adaptation inhered in metalearning are crucial to cross-context learning*

A second implication for the next generation of AI solutions stems from the plasticity and adaptability intrinsic to the metalearning scaffolding that supports morally aware learning in the developing brain. The human capacity to create new as well as to restructure existing neuronal connections, rewiring the brain to alter its execution in response to environmental stimuli or learning activities, may empower young children to spark or suppress plastic episodes in the brain and adapt their explore-exploit strategies in real time to tackle diverse problems. The neuroplasticity encountered in children to help them negotiate the variable-predictable environment dilemma via recursive explorative and exploitative strategy adaptations can serve as a model for integrating plasticity into neural network architectures. Without plastic and adaptable computational models driving AI systems, we cannot achieve synthetic intelligence that is truly versatile and responsible. In the context of a deep neural network (DNN), Lyle et al. (2023) define plasticity as the responsiveness of the network to update model predictions based on new stimuli. The authors contend that network plasticity may be prevented from being attenuated during training by performing layer normalization in deep reinforcement learning (RL) systems. Chalvidal et al. (2022), inspired by biological fast synaptic plasticity, propose a versatile meta-RL neural network that employs a meta-learned synaptic update rule to enable it to self-modify its weights recursively for task rule identification and policy adaptation while implementing diverse cognitive functions. The development of responsible and adaptive AI can be accelerated with the application of fast artificial plasticity in, say, self-learning networks to do more, such as dynamic neural network reorganization that may support adaptive connections across a self-optimizing neural computational model. Prospective edge-AI systems are likely to support not only more plasticity and adaptability but also sparser semantic neural network topologies which comprise elements that discriminatively ascribe greater attention to stronger associations and systematically push weaker correlations in data through a memory-clearing process to mitigate semantic misrepresentations. These systems can also consider making network sparsity a programmatically controllable or autonomously adaptable hyperparameter to better adjust to new contexts. Contextual adaptation is observed in young children—when sparse networks present a disadvantage and denser networks are required instead—who tend to enhance sparse semantic networks through experienced learning and knowledge acquisition (Gualtieri & Finn, 2022). Manipulating sparsity in computational models has been documented in literature. Kim et al. (2016) regulate the parameter sparsity in hidden layers of a deep neural network using 1-norm regularization, while Bennett and Parrado-Hernandez (2006) mention the use of a trade-off parameter to control the level of 1-norm-enforced sparsity in direct kernel pattern classification models like support vector machines. The art and science of crafting pretrained context-aware ML/DL models, which are self-adapting to the degree of sparsity and plasticity required for optimal learning in diverse problem contexts as well as to the scalability demands of expanding learning workloads and complexities, can be refined by mapping the AI infrastructure processes on to the essential learning mechanisms that accord learning versatility to young children. Accordingly, developing adaptive AI solutions with integrated attention and memory-clearing processes can endow artificial learners with the capacity to metalearn efficient episodic control, while an embedded multimodal feedback mechanism with attention can enable metacognitive awareness of both the artificial synaptic states and the action-reward representations in advanced meta-RL agents.

*7.2.3 Strategic tradeoffs constitute an indispensable exemplar for morally aligned artificial neural computing*

A third key inference concerns the tradeoffs that characterize the human neurobiological system. The proposed metalearning framework attests to the capability of young children to generate a range of compromises in their resource-constrained brain by allowing several competing cognitive activities to contend for attentional resources concurrently, thereby advancing cognitive flexibility and performance. Tradeoffs provide the necessary rationale for choice creation as well as strategy selection and discovery. Without exercising the option to swap one position for another between adversarial activities, the human neurobiological system would likely fail to yield new improved strategies and attain optimal outcomes. Some tradeoffs are intrinsic to artificial neural computing driven by ML/DL models and involve harmonizing adversarial goals or aspects of either the embedded models or the overall embedding system. For example, striking a balance between the bias and the variance of the learned computational model represents the foundational compromise in statistical ML (Dar et al., 2021). The bias-variance dilemma, which in this paper denotes a *procedural* tradeoff given its origin in the model training procedure, influences model test performance and generalizability. Tradeoffs, which seek a compromise, for instance, between validity and tractability (Bennett & Parrado-Hernandez, 2006), or accuracy and speed (Arzani et al., 2022), or complexity and a combination of factors including energy utilization and speed (Borrego-Carazo et al., 2020) of the deployed computational models, are classified herein as *functional* due to their fixed purpose-driven emphasis on realizing the high-level provisional goals of the embedding AI systems. In contrast, fine-grained *strategic* tradeoffs effect a deliberate balance between adversarial learning strategies or components to yield sustained cross-context learning improvements. A suitably large combination of strategic compromises directed with a high degree of specificity in the metalearning-enabled developing brain is the cornerstone of impressive cognitive performance evinced by young children. Some benefits of employing synthetic analogues of biological strategic tradeoffs to enhance AI-based model outcomes have been documented in literature. Chalvidal et al. (2022) contend that the learning efficiency of RL systems, charged with identifying the underlying task structure and activating reflexive feedback control, may be improved with artificial learning agents balancing the fast policy update discoveries in the agent policy space against the preservation of explorative performance improvements. Bouchard and Triggs (2004) claim that the trade-off between a discriminative classification strategy and its generative equivalent (using linear interpolation between two corresponding maximization constraints) yields a potentially superior intermediate classifier. Notwithstanding the advantages of the artificial strategic tradeoffs, the optimization routines ensconced in the algorithmic underpinnings perform their function most effectively even if the outcome is to the detriment of salient moral considerations that are outside the domain of the corresponding objective functions. The prospect of augmenting artificial computational architectures by integrating a wide array of strategic tradeoffs inspired by the orchestrated adjustments that the fundamental metalearning components routinely make in the morally grounded developing brain bears immense long-term significance to the future of AI development. These tradeoffs stand to gain greater significance, imaginably assuming a mantle of indispensability, in the presence of resource limitations. Thus, incorporating multiple strategic tradeoffs into resource-constrained AI solutions—while drawing inspiration from the developing brain to deliver increasingly adaptive, morally aligned, reliable, and energy-efficient outcomes—can become the new holy grail of responsible artificial neural computing.

*7.2.4 Strategies for attention, explore-exploit adaptation, feedback monitoring and evaluation, and learning transfer may organically elicit moral responsiveness in metalearning-augmented AI technologies*

Last but not the least, the implication that the impressive learning prowess of young children is intertwined with their moral responsiveness is important. It is especially pertinent to the perspective that moral reasoning and decision-making need not be directly embedded in software blueprints to drive principled AI technologies of the future[17]. In fact, morally grounded outcomes in augmented AI systems (for example, neuromorphic AI incorporating the theorized metalearning components) can potentially emerge as a corollary of allocating attentional control, adapting explore-exploit policy, monitoring and evaluating multimodal feedback, and enabling multifaceted learning transfer (while managing a unified compute-storage infrastructure) across learning tasks in a metalearning context. Even if the genesis of morality is plausibly manifested in metalearning-augmented AI solutions, critics of anthropomorphic connectionism may be unwilling to relinquish human control in AI-guided scenarios. Watson (2019) indicates that the responsibility for moral decision-making is a normative requirement for making judgments in sensitive contexts, and that the human domain expert represents the nucleus of ethical accountability, especially, since the artificial neural networks powering contemporary AI technologies are typically less robust and reliable than their biological counterparts. However, augmenting AI systems with the proper apparatus to promote a greater level of trust in them may eventually become a matter of practical necessity in high-risk decision scenarios. Where AI systems are integrated with the posited metalearning infrastructure, their compliance with a morally oriented development and design roadmap need not be entrusted to humans in the loop or, otherwise, be placed on the proverbial back burner. A simplistic scenario, for example, may involve an edge-AI solution based on a metalearning-driven RL system comprising an artificial learning agent engaged in exploiting previously learned state-action-reward representations tagged in memory and exploring unknown input states in the given learning context. As the RL agent executes an appropriate explore-exploit policy to maximize an engineered reward function that incorporates a designated principled learning incentive, the integrated attention mechanism helps to shape reward dynamics to provide the feedback mechanism with promising behavioral information during training episodes (communicated back to the agent via intra-learning transfers). Consequently, the trained agent picks up moral rules that its biological counterpart generally learns during cognitive development. It is not a giant leap of imagination to envisage a world in which future neuromorphic technologies equipped with adaptive intelligence for deploying the necessary ethical guardrails may well pave the way for responsible AI development. Put simply, humanity will stand to gain more from powering the next generation of AI solutions with computational networks that learn by emulating the brain of young children rather than with synthetic learning algorithms that fail to incorporate the ingenuity and the integrity of developing minds.

In summary, the proposed parsimonious metalearning representation—conceptualized as a tetrad of essential mechanisms operating in tandem within the metacognitive scaffolding—helps to model the mental processes and characterize the underlying principal features that enable young children to generate surprisingly effective learning outcomes across contexts. Clearly, if an ensemble of four complementary enabling mechanisms can reasonably explain the exceptional cross-context learning outcomes that young children spawn, it can obviate the need to construct more complex model substitutes. However, whether the integrated quaternion configuration is a reliable representation of the learning phenomenon generally observed among typically developing preschoolers and early elementary learners is a matter of experimental research. To get such a study underway, a meaningful corpus of empirical data may be obtained from widely deployed technology-mediated interventions. Meanwhile, the posited conceptual framework may shed some light on the question of observed diversity in cross-learner strategy motivation (Berger, 2023). Value judgments about cognitive strategies, which spur learners to act on these approaches, vary across learners likely due to potential differences in their application of attention and memory, resolution of the explore-exploit dilemma, operation of feedback control, and activation of intra and inter-learning transfers. Clearly, the variation in strategy motivation ought to be evident between children with and without learning difficulties or disabilities. This paper, however, makes no attempt to theorize how the four essential enabling mechanisms constituting the proposed metalearning architecture would adapt to the needs of the learning-disabled young children. Research shows that children who experience broad difficulties in learning tend to learn slowly, apply strategies ineffectively, and perform poorly in school (Bosson et al., 2020). However, when exposed to metacognitive training intervention via human mediator-guided prompting, children with learning difficulties show improvement in their metacognitive knowledge awareness and strategy application (Bosson et al., 2020). This result seems to be consistent with the role that the multimodal feedback mechanism performs within the theorized metalearning framework, allowing the attention mechanism to reinforce strategy execution on ineffective learning outcomes via intra-learning transfers. Perkins and Salomon (1992), citing research by Belmont et al. (1982) on intellectually disabled children, postulate that encouraging self-regulation in strategy application training interventions potentially helps children with intellectual disabilities to optimize their application of learned strategy in transfer learning contexts. However, it is unclear whether the composition and functionality of the processes underpinning metacognitive learning would vary significantly between student populations with and without learning disabilities, especially in the presence of advanced (i.e., interactive and adaptive AI) technology-based mediation. Additional research on metalearning infrastructure adaptation for children with learning impairments is needed to understand both the nature and the degree of fine-tuning that the constituent enabling mechanisms may require. Integrating children with learning disabilities into mainstream learning environments and eliciting their active participation in shared learning activities may also require further examination into the underpinnings of metalearning. For example, working memory capacity (an integral constraint on attentional control) is likely to be a critical factor in influencing the effectiveness of collaborative learning (Du et al., 2022). Accordingly, how would the attention-memory dynamics adjust for intellectually disabled children in a collaborative learning landscape? Moreover, this paper extends the notion of transfer learning, having proposed two new transfer paths and advanced the possibility of adaptations within the transfer path space. Therefore, two follow-up questions seem relevant for future metalearning research. First, can the suggested metalearning architecture foster feasible pairwise synthesis of transfer paths to yield new cohesive compound channels (e.g., *free-space low transfer path* for the transfer of reflexive strategies that may degenerate over certain learning contexts) in an experimental setting? Second, which learning transfer paths may be utilized by learning-disabled children for negotiating morally compliant problem-solving challenges across contexts in collaborative learning environments? By formulating a bare-bones metalearning framework to illuminate an irreducible configuration of four fundamental enablers of the inspiring learning competence observed in young children, we hope to motivate and shape additional theoretical and experimental metalearning research. Through this paper, we aspire to set the stage for evidence-based investigations along two dimensions. One, an in-depth examination of how the fundamental metalearning components may contribute to the morally responsive learning successes achieved by young children with learning disabilities. Two, a study of the implications that the outcome of the research along the first dimension may have in helping to influence the development of scalable and responsible self-learning AI systems inspired by the learning-disabled children's neurobiological limitations. If we can learn from analyzing how learning-disabled children use their knowledge about their learning to succeed in disparate problem contexts while comporting with moral rules, we can transfer the lessons learned to more challenging AI-based learning scenarios that necessitate ethical considerations and principled judgments.

## 8. Conclusion

Neuromorphic AI systems emulate the human brain with synthetically orchestrated neural networks, typically requiring copious amounts of training data and time to learn highly accurate, albeit morally agnostic, generalizations across multiple domains. In contrast, the developing brain possesses the competence to rapidly infer broad patterns from a solitary event, applying knowledge learned in one context to another. This paper associates the construct of self-regulated learning (SRL) with a unique minimalistic configuration of cognitive processes (attention, explore-exploit tradeoff, feedback, and transfer) to introduce a novel metalearning framework rooted in developmental psychology and neuroscience theories. The proposed framework not only serves to elucidate the remarkable and morally congruent learning among young children but also forms the basis for articulating implications for a morally grounded AI blueprint. Accordingly, this article details the four framework constituents and reveals their interrelationships that support awareness and regulation of learning. A suitable configuration of process as well as strategy implementations within the posited metalearning architecture is offered as a conceivable path to achieving learning competence. Further, salient insights applicable to brain-inspired AI systems are inferred by recognizing the parallels between the core cognitive processes executed by young neurobiological learning agents and the mechanisms operated by their neuromorphic counterparts. In this context, we suggest equipping future neuromorphic systems with our innovative metalearning framework to empower the embedded AI to deliver responsible learning and problem-solving outcomes akin to those engendered by developing minds. Furthermore, we extend the theory of learning transfer in this paper to accommodate both intra- and inter-learning transfers along two new transfer path constructions that may be particularly germane to morally guided decision-making by artificial neural networks in complex scenarios.

**Notes**

[1] In the paper, "moral" subsumes "ethical", as well as the associated derivative terms, without making an express distinction among the underlying concepts. Herein, moral and ethical behaviors or outcomes are considered within the same construct of morality guiding human actions that abstain from causing harm (even rejecting the principle of harming a few to prevent harming a multitude).

[2] For simplicity, this paper assumes that each problem context or domain is associated with an individual problem-solving task to resolve. Therefore, "task learning" and "learning in the given context" have the same connotation in this work.

[3] Zhang and Zhang (2019) define the concept of self-regulated learning as "learning how to learn".

[4] See "hoped-for transfer" in Perkins & Salomon (1992).

[5] The notion of "memory clearing" in this paper is used in relation to strategies, procedures, functions, or activities that work to mitigate cognitive load by reducing the memory footprint for neurobiological computing. Essentially, it is the subset of human memory management which includes the cognitive processes of information retention and retrieval. Since memory clearing facilitates retention and retrieval, this paper employs the narrower concept as a proxy for its more comprehensive counterpart.

[6] The inhibition of uninformative or irrelevant sensory stimuli is known as "sensory gating" (see Watson, 2019) in psychological parlance.

[7] See Gualtieri & Finn (2022).

[8] A metadata tag is a construct introduced in this work to denote information about information in memory. It represents a key information fragment or label with a minimal memory footprint used for tagging of information—including data associated with actions, behaviors, events, feelings, and objects—traced in memory. In as much as the metadata tags accelerate retrieval of the associated memory-intensive perceptual details, these labels are analogous to memory-efficient stimuli or triggers used as memory cues. In this paper, metadata tags are theorized as being integral to the memory clearing strategy executed by young children for the tracking, reactivation, retrieval, and removal of full memory traces. A comprehensive discussion on memory metadata tags is reserved for future research.

[9] In contrast, the discovery-heuristics component of the SCADS model in Shrager and Siegler (1998) operates on working memory traces of strategy operations. Although their model is proposed as an isomorphic representation of young children's cognitive operations (i.e., strategy selection, execution, and discovery) involved in solving arithmetic addition problems, these children may be employing a more efficient neurobiological process than the full memory trace they are believed to create per strategy execution instance for retrieval purposes. The extremely transient nature of traces in working storage and the unwieldy constraints on rehearsal rates at which these traces can be optimally refreshed (Towse et al., 2000) may be circumvented by the use of memory metadata tagging.

[10] To be suitable, the implemented strategy must address the given problem accurately, subject to the learner's satisfaction. Therefore, the right strategy may not be suitable if the learner is not satisfied with the strategy's outcome (e.g., sluggish execution, unacceptable resource consumption, or unexplored superior options). Rather than dismissing such a strategy as unsound or ineffective, it may be incorporated as the basis for synthesizing a new, potentially high-performing strategy within the framework.

[11] See "catastrophic forgetting or interference" in Singh (2023).

[12] Hereafter, the terms—"NK-B4P" (new knowledge, behaviors, perceptions, preferences, principles, and proficiencies), "experience", "knowledge", and "learning"—are used interchangeably unless a distinction or qualification is expressly stated.

[13] See Austerweil et al. (2019).
[14] See "near" and "far" transfers in Perkins and Salomon (1992).
[15] See "affordances" in Perkins and Salomon (1992).
[16] To be consistent with other proposed transfer mechanisms, this paper prefers to use "path" instead of "road", the original term employed by Perkins and Salomon (1992) to describe their "psychological paths by which transfer occurs".
[17] This view diverges from the notion of encoding ethics expressly into the programming of machine learning algorithms that aim to overcome social challenges, including privacy and bias, as described in Kearns and Roth (2019).

Huang, Q. (2022). Weight-quantized squeezenet for resource-constrained robot vacuums for indoor obstacle classification. *AI*, *3*(1), 180-193.

Kang, C. Y., Duncan, G. J., Clements, D. H., Sarama, J., & Bailey, D. H. (2019). The roles of transfer of learning and forgetting in the persistence and fadeout of early childhood mathematics interventions. *Journal of Educational Psychology*, *111*(4), 590–603.

Kearns, M., & Roth, A. (2019). *The ethical algorithm: The science of socially aware algorithm design*. Oxford University Press.

Khasawneh, M., Alkhawaldeh, M., & Al-Khasawneh, F. (2020). The Level of Metacognitive Thinking Among Students with Learning Disabilities. *International Journal of English Linguistics*, *10*(5).

Kim, J., Calhoun, V. D., Shim, E., & Lee, J. H. (2016). Deep neural network with weight sparsity control and pre-training extracts hierarchical features and enhances classification performance: Evidence from whole-brain resting-state functional connectivity patterns of schizophrenia. *Neuroimage*, *124*, 127-146.

Kim, S. J., Kim, S. B., & Jang, H. W. (2021). Competing memristors for brain-inspired computing. *Iscience*, *24*(1), https://doi.org/10.1016/j.isci.2020.101889.

Le, V. N., Schaack, D., Neishi, K., Hernandez, M. W., & Blank, R. (2019). Advanced content coverage at kindergarten: Are there trade-offs between academic achievement and social-emotional skills? *American Educational Research Journal*, *56*(4), 1254-1280.

Lyle, C., Zheng, Z., Nikishin, E., Pires, B. A., Pascanu, R., & Dabney, W. (2023). Understanding plasticity in neural networks. *arXiv preprint arXiv:2303.01486*.

Maass, K., Zehetmeier, S., Weihberger, A., & Flößer, K. (2023). Analysing mathematical modelling tasks in light of citizenship education using the COVID-19 pandemic as a case study. *ZDM–Mathematics Education*, *55*(1), 133-145.

Mehmood, F., Ahmad, S., & Whangbo, T. K. (2022, February). Object detection based on deep learning techniques in resource-constrained environment for healthcare industry. In *2022 International Conference on Electronics, Information, and Communication (ICEIC)* (pp. 1-5). IEEE.

Mehonic, A., & Kenyon, A. J. (2022). Brain-inspired computing needs a master plan. *Nature*, *604*(7905), 255-260.

Niv, Y., Daniel, R., Geana, A., Gershman, S. J., Leong, Y. C., Radulescu, A., & Wilson, R. C. (2015). Reinforcement learning in multidimensional environments relies on attention mechanisms. *Journal of Neuroscience*, *35*(21), 8145-8157.

Perkins, D. N., & Salomon, G. (1992). Transfer of learning. *International Encyclopedia of Education*, *2*, 6452-6457.

Prather, J., Becker, B. A., Craig, M., Denny, P., Loksa, D., & Margulieux, L. (2020, August). What do we think we think we are doing? Metacognition and self-regulation in programming. In *Proceedings of the 2020 ACM Conference on International Computing Education Research* (pp. 2-13).

Rakoczy, H., Hamann, K., Warneken, F., & Tomasello, M. (2010). Bigger knows better: Young children selectively learn rule games from adults rather than from peers. *British Journal of Developmental Psychology*, *28*(4), 785-798.

Rule, J. S., Tenenbaum, J. B., & Piantadosi, S. T. (2020). The child as hacker. *Trends in Cognitive Sciences*, *24*(11), 900-915.

Schraw, G. (1998). Promoting general metacognitive awareness. *Instructional Science*, *26*(1-2), 113-125.

Schraw, G., Crippen, K. J., & Hartley, K. (2006). Promoting self-regulation in science education: Metacognition as part of a broader perspective on learning. *Research in Science Education*, *36*, 111-139.
Singh (Sep. 2023)     26